\documentclass[twocolumn]{aastex631}

\usepackage{siunitx}
\usepackage{amsmath}
\usepackage{natbib}
\usepackage[normalem]{ulem}
\hypersetup{}

\newcommand{\eo}{\textcolor{cyan}{}\bgroup\markoverwith{\textcolor{cyan}{\rule[.5ex]{2pt}{2.5pt}}}\ULon}

\begin{document}

\title{Evidence for an Outer Component in the Continuum Reverberation Mapping of Active Galactic Nuclei}

\email{yzjiang@stu.pku.edu.cn; wuxb@pku.edu.cn}
\author[0009-0001-5163-5781]{Yuanzhe Jiang}
\affiliation{Department of Astronomy, School of Physics, Peking University, Beijing 100871, China}
\affiliation{Kavli Institute for Astronomy and Astrophysics, Peking University, Beijing 100871, China}

\author[0000-0002-7350-6913]{Xue-Bing Wu}
\affiliation{Department of Astronomy, School of Physics, Peking University, Beijing 100871, China}
\affiliation{Kavli Institute for Astronomy and Astrophysics, Peking University, Beijing 100871, China}

\author[0000-0003-0827-2273]{Qinchun Ma}
\affiliation{Department of Astronomy, School of Physics, Peking University, Beijing 100871, China}
\affiliation{Kavli Institute for Astronomy and Astrophysics, Peking University, Beijing 100871, China}

\author{Huapeng Gu}
\affiliation{Department of Astronomy, School of Physics, Peking University, Beijing 100871, China}
\affiliation{Kavli Institute for Astronomy and Astrophysics, Peking University, Beijing 100871, China}

\author{Yuhan Wen}
\affiliation{Department of Astronomy, School of Physics, Peking University, Beijing 100871, China}
\affiliation{Kavli Institute for Astronomy and Astrophysics, Peking University, Beijing 100871, China}

\begin{abstract}

The continuum reverberation mapping is widely used in studying accretion disk of active galactic nuclei (AGN). While some indirect evidence and simulations indicated that the diffuse continuum, especially the strong Balmer continuum from the broad line region (BLR), may contribute to the continuum in the u/U band. Here, we present direct evidence for this contribution. In this work, we apply the ICCF-Cut method to continuum reverberation mapping to extract the possible diffuse continuum light curves of 6 AGNs with high cadence, high quality and multi-band observations. We find the existence of an outer component out of the accretion disk for each of 6 AGNs in the Swift U band. Meanwhile, similar results can be derived by JAVELIN Photometric Reverberation Mapping Model for 4 of them. The lags of the outer components are consistent with the predicted Balmer continuum lags, which are about half of the H$\beta$ lag values. Our result directly reinforces that an outer component, especially the Balmer continuum in the rest-frame u/U band, can contribute significantly to the continuum reverberation lags of AGNs.

\end{abstract}


\section{Introduction}
Continuum reverberation mapping (CRM hereafter) is a powerful method to probe the unresolved structure and sizes of continuum emitting regions in AGNs. The basic principle is that the UV/optical continuum variability in AGNs is the reprocessing of a central point-like emitting source according to the lamp-post model \citep{1991ApJ...371..541K, 2007MNRAS.380..669C}. Within the framework of this model, the variations in different wavelengths will correlate to each other but have time delays corresponding to the distance between those different continuum emitting areas. Recent observations support this assumption. The continuum lags are found in targets with high quality, high cadence and multi-band monitoring \citep[e.g.,][]{2015ApJ...806..129E, 2016ApJ...821...56F, 2017ApJ...840...41E, 2018ApJ...854..107F, 2018MNRAS.480.2881M, 2019ApJ...870..123E, 2020ApJ...896....1C, 2020MNRAS.498.5399H, 2021ApJ...922..151K, 2021MNRAS.504.4337V}, as well as in some larger samples with light curves with less cadence and a few bands \citep[e.g.,][]{2017ApJ...836..186J, 2018ApJ...862..123M,
2020ApJS..246...16Y,
2022ApJ...940...20G, 2022MNRAS.511.3005J, 2022ApJ...926..225H}. 

Previously, continuum reverberation mapping was used to measure the size of accretion disk around the super-massive black hole. Since the main contribution in a broad UV/optical band was thought to be from the accretion disk, the broad-band photometric reverberation mapping is often applied in this measurement. According to the widely used disk model, called the standard thin disk or Shakura-Sunyaev disk  \citep{1973A&A....24..337S}, the accretion disk is assumed to be geometrically thin, optically thick, and can emit thermal continuum radiation across the UV-optical wavelength. This disk model predicts that the variations of the cooler outer disk are delayed from the hotter inner disk due to the different light travel time from the central source, and the lags between different bands follow the relation $\tau \propto \lambda ^{4/3}$ \citep[e.g.,][]{2007MNRAS.380..669C}. This similar relation has been confirmed by many CRM works we mentioned above \citep[e.g.,][]{2016ApJ...821...56F, 2020ApJ...896....1C, 2017ApJ...836..186J}. 

Although from observations, light curves in different bands are correlated with each other and the lags are proportional to $\lambda^{4/3}$  approximately, the disk sizes derived from the CRM, however, are usually 2-3 times larger than the standard thin disk model predictions. Moreover, targets with high quality multi-band observations always show a u/U band excess in their lag-spectrum, i.e. the lag in the u/U band is larger than that in other bands around the u/U band \citep[e.g.,][]{2016ApJ...821...56F, 2018MNRAS.480.2881M, 2020ApJ...896....1C}. The u/U band excess contradicts the relation that $\tau$ increases with $\lambda$. Therefore, some further explanations are needed.

A possible explanation has been proposed and discussed for a long time. It claims that the diffuse continuum from BLR is responsible for this phenomenon \citep[see some discussions on CRM in papers cited above and some simulation evaluations in][]{2001ApJ...553..695K, 2018MNRAS.481..533L, 2019MNRAS.489.5284K, 2020MNRAS.494.1611N}. Diffuse continuum is caused by free-free or free-bound emissions from the BLR, which is located out of the accretion disk in AGN. Therefore, continuum light curves encompass contributions not only from accretion disk but also from some components in the outer region, and the observed lags will be larger than the predicted lags based solely on the disk model. This effect will be more evident in the u/U band because of the strong Balmer continuum and Balmer jump ($3647 \rm{\mathring{A}}$) in the BLR. The contribution of Balmer continuum also has observational features in AGN's spectrum and spectral energy distribution (SED). In some spectral decomposition and SED fitting works \citep[e.g.,][]{2016A&A...588A.139M}, Balmer continuum is thought to be responsible for the small blue jump, which can contribute a lot to the continuum emission in the u/U band. Consequently, Balmer continuum from the BLR may increase the lag between u/U band and other bands significantly, which could explain the excess of lags in the u/U band potentially.

Inspired by these observational effects, some simulation works have been done to evaluate the contribution of diffuse continuum to the continuum lags \citep[e.g.,][]{2018MNRAS.481..533L,2019MNRAS.489.5284K,2020MNRAS.494.1611N,2022MNRAS.509.2637N}. They simulated the AGN ionization based on some assumptions of AGN models and found that the diffuse continuum can contribute a lot to the total lags by simulation results. They claimed that the contribution of diffuse continuum lags may be dominated because the diffuse continuum lag is much lager than the disk lags, even if its emission fraction is lower than that of the disk. Moreover, \citet{2022MNRAS.509.2637N} found that the total modeled lag is consistent with observations, if assuming a typical Balmer continuum lag $0.5\tau(\rm{H\beta})$, being half of the $\rm{H\beta}$ lag, determined by CLOUDY in \citet{2020MNRAS.494.1611N} and the continuum lag to be the weighted average of the disk lag and diffuse continuum lag by their own flux values. A new lag-luminosity relation has also been discovered, which shows the continuum lags following the $R \sim L^{0.5}$ approximately like the $R-L$ relation of BLR because the diffuse continuum is dominated in some bands \citep{2022MNRAS.509.2637N,2022ApJ...940...20G,2023ApJ...948L..23W}.

Those investigations claims that the diffuse continuum can contribute to the continuum reverberation mapping based on the fact that the diffuse continuum flux can not be neglected and it is emitted from the BLR out of the accretion disk. However, it can not ensure that the diffuse continuum can reverberate to the disk emissions and provide a large observed lag, so the evidence mentioned above is indirect. It is important to provide more direct evidence to show if there are outer components beyond accretion disks in the broad-band photometric light curves that have good correlations with the inner disk variability and provide us with longer lags than the lags predicted by disk models. In this paper, we adopt a method called ICCF-Cut for continuum reverberation mapping \citep[see the previous ICCF-Cut application to line emission in][]{2023ApJ...949...22M}. By applying this method, we decompose the light curve of the potential diffuse continuum from the Swift U band in 6 AGNs with high quality, high cadence and multi-band monitoring light curves. The method is based on the model of BLR structure and its emission properties. We further investigate the correlation and the lag between the resulting light curve and the inner continuum emission. To increase the reliability of ICCF-Cut results, the JAVELIN Photometric Reverberation Mapping Model (Pmap Model hereafter) is also applied to those targets. This method can provide another estimation of the lag of this outer component. 

This paper is arranged as follows. In Section 2, we describe the ICCF-Cut method for continuum reverberation mapping, decomposing the observational light curves based on some assumptions, and give a brief introduction of JAVELIN Pmap Model. In Section 3, we describe the sample selection and then apply the ICCF-Cut and JAVELIN Pmap Model to those targets to calculate the lags of different components in Section 4. Finally, we discuss our results in Section 5 and summarize our conclusions in Section 6.

\section{Methodology}

\subsection{ICCF-Cut Method for Continuum Reverberation Mapping}

First, we provide a brief introduction to the ICCF-Cut method. It was developed by \citet{2023ApJ...949...22M} to measure the emission line lags from the broad-band photometric light curves originally. The method provides a possibility to estimate the broad line region sizes and black hole masses of a large sample of AGNs in the large multi-epoch high cadence photometric surveys. It can remove the continuum component from a band with a blend of continuum and strong line emission. We call this band as ``line band", and the band with almost no line contamination as ``continuum band". We employ the continuum band to estimate the continuum flux level in the line band using the single-epoch spectrum data, from which we can infer the fraction of the continuum component. After knowing about the light curve shape and mean flux of the continuum component in the line band, we can remove it from the line band and get a resulting light curve called the ``cut" light curve which only has line emission in the ideal situation. Finally, we run interpolated cross-correlation function \citep[ICCF; e.g.,][]{1987ApJS...65....1G} to calculate the emission line lag. This method has recovered $\rm{H \alpha}$ lags for four Syfert 1 galaxies (MCG+089 11-011, NGC 2617, 3C 120 and NGC 5548) successfully \citep{2023ApJ...949...22M}. 

As described above, ICCF-Cut is a method that extracts the component embedded in the broad-band photometric light curve. After that, the lag of this component to the continuum band can be estimated by ICCF. This is quite similar to the situation we care about, where a strong diffuse continuum instead of the line emission is embedded in a certain band. Therefore, we will adjust the ICCF-Cut method to the CRM by changing the light curve in the ``continuum band" to the light curve with pure disk component and decomposing the embedded light curve based on the diffuse continuum model, and to see if the extracted light curves for possible diffuse continuum have a good correlation with the light curves for the accretion disk and have larger lags than previous observed continuum lags.

\subsubsection{Disk Component}

To investigate if any outer component can contribute to the continuum lag, we should remove the disk component from the total light curve. To realize this goal, we must know the light curve shape and mean flux of the disk continuum in the broad band.

Firstly, we aim to understand the composition of the continuum spectrum by simulations. We follow \citet{2022MNRAS.509.2637N} and simulate the AGN ionization by CLOUDY \citep{2017RMxAA..53..385F}. The detailed discussion can be found in \citet{2020MNRAS.494.1611N,2022MNRAS.509.2637N}. For completeness, we introduce the simulation strategy briefly. In those papers, the BLR cloud structure is assumed to be determined by the radiation pressure of the central source, which is called radiation pressure confined (RPC) clouds \citep{2002ApJ...572..753D,2014MNRAS.438..604B}. The SED used here refers to AD1 in \citet{2020MNRAS.494.1611N}. It adopts an optically thick and geometrically thin disk SED calculated by \citet{2012MNRAS.426..656S}, combined with an X-ray power-law continuum. Given the SED, BLR cloud model and luminosity of the AGN, we can derive the AGN ionization by CLOUDY. This provides the luminosity of the simulated disk and diffuse continuum. Therefore, we can estimate the fraction of the diffuse continuum at different wavelengths. 

We noticed that there are still some other BLR models which are very different from each other. \citet{2022MNRAS.509.2637N} claims that the wavelength dependencies of predicted lags due to the diffuse continuum emission are similar. We also find that the diffuse continuum ratios of NGC 5548 predicted by different models in the Swift U band are all about 40 \%. Our discussion below shows that the diffuse continuum ratio is what we care about most. Additionally, as mentioned in Section 1, the lag spectrum predicted by \citet{2022MNRAS.509.2637N} is similar to the observations, and a new R-L relation of continuum lags has been proposed and confirmed in some works \citep[e.g.,][]{2023ApJ...948L..23W}. Those investigations demonstrate that the RPC model is plausible for the ICCF-Cut method of continuum reverberation mapping,  so we will just use the RPC model. We will discuss the influence of different diffuse continuum ratios in Section 5. Further comparison among different models is beyond the scope of this paper.

\begin{figure}[!ht]
  \includegraphics[width=1.0\linewidth]{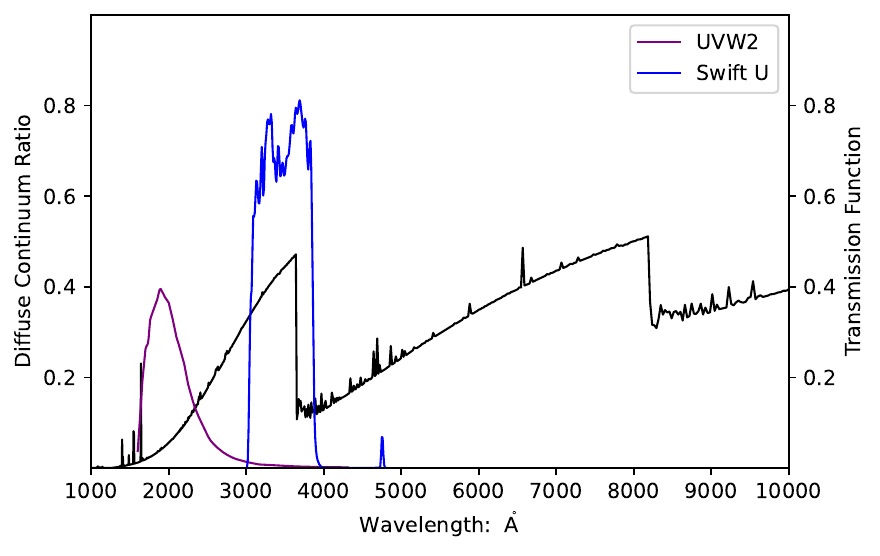}
  \caption{The diffuse continuum ratio spectrum of NGC 5548 and the transmission functions of the Swift UVW2 and U bands. The black line shows the wavelength dependencies of the diffuse continuum ratio predicted by the CLOUDY simulation for a standard RPC cloud model.} 
  \label{dc_ratio}
\end{figure}

Figure \ref{dc_ratio} shows an example of simulated diffuse continuum ratio spectrum in NGC 5548, and other targets have similar diffuse continuum ratio spectra. The diffuse continuum plays a negligible role in the bluer band ($<2500 \rm{\mathring{A}}$) and significantly impacts the redder band. The ratio is extremely high when it reaches the Balmer jump ($3647 \rm{\mathring{A}}$) and the Paschen jump ($8206\rm{\mathring{A}}$). Since there is observational evidence of Balmer continuum contribution, we will focus on trying to derive an outer component in the u/U band. We will assume the outer component to be the diffuse continuum in the following discussions, and the validity of this assumption will be evaluated in Section 4.

The total U band light curves can be written as
\begin{eqnarray}
\label{eq_cut_u}
L_{U}(t) = L_{U,disk}(t) + L_{U,dc}(t),
\end{eqnarray}
where $L_{U}(t),L_{U,disk}(t)$ and $L_{U,dc}(t)$ represent the total, disk and diffuse continuum light curves. As mentioned above, the main idea of ICCF-Cut is to get rid of the contaminated component from the total light curve. Here, we will remove $L_{U,disk}$, so we need to know its light curve shape and mean flux.  

In ICCF-Cut, we need a driving light curve with a pure component to get the AGN variability, which is often thought to be the light curve of the inner region and drive the variability in the other region based on the assumption of the reverberation mapping. In Figure \ref{dc_ratio}, the purple and blue curves show the transmission functions of the Swift UVW2 band and U band, respectively. The diffuse continuum ratio in the UVW2 band is far smaller than that in the Swift U band. The specific ratio in Swift U band can be found in Table \ref{cut_result}. Some targets have a lower diffuse continuum ratio because we set the BLR covering factor to half of the other targets. In that way, the modeled lags can fit well with the observed lag spectrum, measured by the ICCF between the observed UVW2 and Swift U band light curves (see discussions in \citet{2022MNRAS.509.2637N} and Section 5). Therefore, we assume that the UVW2 band can be seen as a pure disk component band compared to other bands at longer wavelengths. Under this assumption, we take the UVW2 band light curve as the driving light curve, which can be written as
\begin{eqnarray}
\label{eq_cut_w2}
L_{W2}(t) = L_{W2,disk}(t),
\end{eqnarray}
where $L_{W2}$ and $ L_{W2,disk}$ represent the total and disk light curve respectively.

The variability in the Swift U band will lag behind the variability in the UVW2 band, so we should shift the UVW2 light curves by a lag $\tau_{disk}$ to transfer the disk continuum. $\tau_{disk}$ represents the lag between disk components in the UVW2 band and the Swift U band. Like Equation (12) in \citet{2016ApJ...821...56F}, the disk size R given by the standard disk model is
\begin{eqnarray}
\label{eq_ss_disk}
R = (X\frac{k\lambda_0}{hc})^{4/3} [(\frac{GM}{8 \pi \sigma})(\frac{L_{Edd}}{\eta c^2})(3+ \kappa) \dot{m}_{edd}]^{1/3},
\end{eqnarray}
where $X$ is a multiplicative scaling factor of the order of unity that accounts for the systematic issues in converting the annulus temperature $T$ to wavelength $\lambda$ at a characteristic radius $R$, $L_{Edd}$ is the Eddington luminosity, $\eta$ is the radiative efficiency in converting mass into energy, $\kappa$ is the local ratio of external to internal heating, assumed to be constant with radius, and the Eddington ratio $\dot{m}_{edd}$. We choose $X=2.49$, $\eta = 0.1$ and $\kappa=0$ as in most of other works \citep[e.g.,][]{2016ApJ...821...56F,2017ApJ...840...41E} and adopt the parameters for 6 AGNs listed in Table \ref{prop} to calculate the disk size. Therefore, $\tau_{disk}$ can be written as $(R_{U} - R_{UVW2})/c$.

The mean flux is predicted according to CLOUDY simulations. We used CLOUDY to estimate the disk continuum proportion $p_{disk}=1-p_{dc}=1-{L_{U,dc}}/{L_{U}}$ in the Swift U band. Then we define a simple scaling parameter $\alpha$ to convert the disk flux:
\begin{equation}
\begin{split}
\alpha &= \frac{L_{U,disk}(t)}{L_{W2,disk}(t-\tau_{disk})} \\ 
&= \frac{L_{U,disk}(t)}{L_{U}(t)} \cdot \frac{L_U (t)}{L_{W2}(t-\tau_{disk})} \\
&\approx (1-p_{dc}) \times {\rm{Median}}[\frac{L_{U}(t)}{L_{W2}(t-\tau_{disk})}].
\end{split}
\label{eq_trans_cut}
\end{equation}
The first term is determined by the CLOUDY simulations, and the second term is decided directly by observations in the Swift U and UVW2 bands. According to Equations (\ref{eq_cut_u}), (\ref{eq_cut_w2}) and (\ref{eq_trans_cut}), we can derive the diffuse continuum light curve in Swift U band
\begin{equation}
\begin{split}
    L_{U,dc}(t) &= L_{U}(t) - \alpha L_{W2,disk}(t-\tau_{disk})\\
    &\approx L_{U}(t) - \alpha L_{W2}(t-\tau_{disk}).
\end{split}
\label{eq_dc}
\end{equation}

\subsubsection{Host Galaxy}

We should also consider the contribution of the host galaxy. The host galaxy can contribute a constant flux to the total flux in each band, which has little influences on both the intrinsic variation of light curves and the lag measurements in the broad-band photometric reverberation mapping. So, only for a few targets, the host flux is estimated carefully in the past continuum reverberation mapping works. However, when we estimate the scaling factor $\alpha$, we need to use the ratio of two fluxes after the host galaxy contribution has been excluded. The diffuse continuum ratio given by CLOUDY is the ratio of pure AGN contribution without the host galaxy contamination. Thus, we need to remove the host galaxy contribution first.

Only NGC 5548 in our sample (see the sample selection in Section 3) has sophisticated host galaxy flux measurement by image-subtraction method \citep{2016ApJ...821...56F}. For other targets, we use another method to estimate the host galaxy flux from the published data.

\begin{figure}[!ht]
  \includegraphics[width=1.0\linewidth]{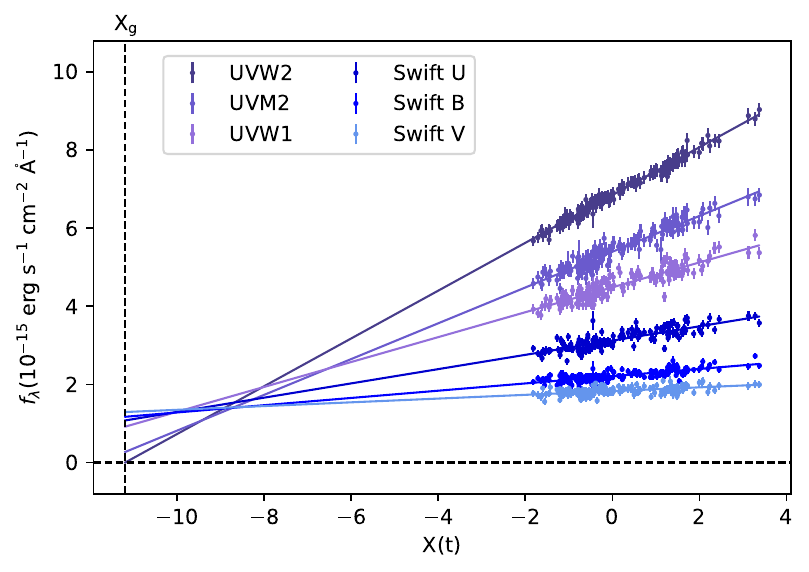}
  \caption{The flux-flux diagram of Mrk 142 for light curves in Swift bands. The data is provided by \citet{2020ApJ...896....1C}. Our results are similar to theirs, but we do not show other optical bands because we only care about the Swift U band. Here, $X_g$  represents the value of $X(t)$ where $f_{\lambda}=0$ for the UVW2 band, which means that the flux-flux relations for other bands at $X_g$ give estimations of the host galaxy contribution.} 
  \label{f-f}
\end{figure}

A frequently-used method to estimate host galaxy flux is the flux-flux analysis \citep[e.g.,][]{2007MNRAS.380..669C, 2017ApJ...835...65S, 2018MNRAS.480.2881M, 2020ApJ...896....1C}. It fits the light curve with a simple linear model
\begin{eqnarray}
\label{eq_ff}
    f_{\lambda}(\lambda,t) = A_{\lambda}(\lambda) + R_{\lambda}(\lambda)X(t),
\end{eqnarray}
where $X(t)$ is a dimensionless light curve with a mean of 0 and standard deviation of 1, $A_{\lambda}(\lambda)$ is the mean of the spectrum and $R_{\lambda}(\lambda)$ is the rms spectrum. If we further assume that the contribution of the host galaxy in the bluest band is negligible, we can derive the host galaxy flux in other bands where $f_{\lambda}=0$ in the bluest UVW2 band. In Figure \ref{f-f}, we plot the flux-flux analysis for Mrk 142, and a good linear response can be seen for all the bands, which is consistent with the flux-flux analysis of Mrk 142 in \citet{2020ApJ...896....1C}. However, this method may cause some uncertainties. As in \citet{2017ApJ...835...65S}, we also found that the host flux measured by flux-flux analysis is a little bit higher than that give by the image-subtraction for NGC 5548. The host flux of NGC 5548 in the Swift U band given by image-subtraction is $\rm{1.22 \times 10^{-15} \ erg \, cm^{-2} \, s^{-1} \, \mathring{A}^{-1}}$, while our result given by the flux-flux analysis is $\rm{5.56 \times 10^{-15} \ erg \, cm^{-2} \, s^{-1} \, \mathring{A}^{-1}}$. By using the flux-flux analysis result, the mean flux without the host galaxy's contribution would decrease about 15\%. The impact about the overestimation of host flux on the lag measurement will be discussed in Section 5.

We do not adopt the flux variation gradient (FVG) used in the previous ICCF-Cut paper \citep{2023ApJ...949...22M} for two reasons: 1) FVG method assumes a good linear relation between fluxes in two bands but the linear relation is not always good for every target in our sample; 2) we have multi-band light curves including UV bands while \citet{2023ApJ...949...22M} only had light curves of \textit{gri} bands, so it is plausible for us to process the flux-flux analysis. With \textit{gri} bands only, we cannot employ such approach. 

\subsubsection{ICCF}

We use the ICCF method to measure the correlation and lag between two bands. ICCF \citep[][]{1987ApJS...65....1G, 1994PASP..106..879W, 1998PASP..110..660P} is a frequently-used lag measurement. The method calculates the cross-correlation function (CCF) to determine the correlation between two light curves. To estimate the lag errors, it uses Monte Carlo simulations to perform the flux randomization and random subset sampling \citep[FR/RSS; more details in][]{1998PASP..110..660P}. Finally, it yields a distribution of centroid lags $\tau_{cen}$ called the Centroid Cross-Correlation Distribution (CCCD). One of the advantages of the ICCF method is that it does not rely on specific AGN models and transfer functions compared to other frequently-used methods, such as JAVELIN \citep[][]{2011ApJ...735...80Z} and CREAM \citep{2016MNRAS.456.1960S}. 

\subsection{JAVELIN Pmap Model}

The basic principle of JAVELIN (Just Another Vehicle for Estimating Lags In Nuclei) is to interpolate the light curves according to the damped random walk (DRW) model and calculate the maximum likelihood to determine lags and relevant parameters by using MCMC methods \citep{2013ApJ...765..106Z, 2011ApJ...735...80Z}. The Pmap Model of JAVELIN is designed to derive emission line lags from pure photometric data \citep{2016ApJ...819..122Z}. It assumes that light curves in the continuum band and line band are $f_c = c(t) + u_{c} $ and $f_l = \alpha \cdot c(t) + l(t) + u_l$. Here, $u_c$ and $u_l$ are constant, and, therefore, mainly represent the contamination of the host galaxy. $c(t)$ and $l(t)$ represent the variability of continuum and line emission in each band, and $\alpha$ shows the ratio of continuum variability in two bands. This model transfers the continuum emission linearly without any lag because the continuum lag is minimal compared to the emission line lag. It converts the continuum variability to line variability by a top-hat transfer function centered on the time lag $\tau$ with the width $\omega$ and the amplitude $A$,
\begin{eqnarray}
l(t) = \int{\Psi(t-t')c(t)dt'}
\end{eqnarray}
\begin{eqnarray}
    \Psi(t) = \frac{A}{\omega} \ \ for \ \ \tau-\omega/2 \leq t \leq \tau + \omega/2
\end{eqnarray}
This light curve model is very similar to the model used in the ICCF-Cut method. We can take the UVW2 band as the continuum band and the Swift U band as the line band. Then, $c(t)$ represents the pure disk variability in the UVW2 band, $l(t)$ represents the diffuse continuum variability in the Swift U band, and $\alpha$ transfers the disk variability between two bands. To remain consistent with the ICCF-Cut, we add the standard thin disk model lag to the disk continuum in the UVW2 band. Thus, we have $c_t = c_{t'}(t'+\tau_{disk})$. We have also run the Pmap Model without adding disk lags, and the result has almost no difference. This is because the disk lag is minimal compared to the outer component lag, and we add the disk lag just for consistency. 

As a result, the Pmap Model is a parallel method to check some parameters of the ICCF-Cut without the complicated modeling of BLR. It can give the posterior distribution of $\tau$, which may provide another estimation of the diffuse continuum lag. The median and $1\sigma$ uncertainties are determined to give the final estimation of these parameters. However, since JAVELIN has many parameters, the distribution of its parameters may not converge to give a robust estimation. In addition, this method assumes a specific model of the AGN variability and a simple transfer function, which is much more model-dependent than the ICCF method. So, we will take ICCF-Cut results as our final results and use the JAVELIN Pmap Model to assess the ICCF-Cut results.

\section{Sample}

We select 6 AGNs with high-quality, densely sampled and multi-wavelength broad-band photometric observations that have published calibrated data. The most important factor is that all these targets have Swift UVOT data. Because we need the UVW2 band with very strong disk contribution and little contamination from other components to be the driving band and use it as the pure disk continuum. Also, the u/U band excess should be observed because it is one of the indirect evidence of the Balmer continuum. These 6 targets are: NGC 5548 \citep{2015ApJ...806..129E, 2016ApJ...821...56F}, NGC 2617 \citep{2018ApJ...854..107F}, NGC 4151 \citep{2019ApJ...870..123E}, Mrk 142 \citep{2020ApJ...896....1C}, NGC 4593 \citep{2019ApJ...870..123E} and Mrk 509 \citep{2019ApJ...870..123E}. Their physical properties can be found in those reference papers and the \href{http://www.astro.gsu.edu/AGNmass/}{AGN Black Hole Mass Database} \citep{2015PASP..127...67B}. We include their properties in Table \ref{prop}. The original lag measurements between the UVW2 and Swift U band are shown in column (6) of Table \ref{cut_result}. 

NGC 4593 also has light curves in many different wavelengths chosen from spectral monitoring. However, we only focus on UVW2 and Swift U bands, so we just use the light curves in the UVOT band from \citet{2019ApJ...870..123E}. More broad-band photometric light curves are also included in  \citet{2017ApJ...840...41E}, but they are almost in the same observation season, and the lags are pretty similar. 

Fairall 9 is also a target that has available multi-wavelength observations, but it is reported to contain a slowly varying component with an opposite lag to the reverberation signal \citep{2020MNRAS.498.5399H}. As described in Section 2, we only assume two variable components: disk continuum and diffuse continuum. Though there are some explanations for this issue and a detrending process has been proposed, we note that the physical nature is still not clear, and the detrending process is arbitrary. We don't want to make thing more complicated, so Fairall 9 is not included in our sample.

\begin{deluxetable}{ccccc}[!ht]
\centering
\tablecaption{Properties of the 6 AGNs in our sample}
\tablehead{
\colhead{Name} & \colhead{Redshift} & \colhead{$\log{M_{\rm{BH}}/\rm{M_{\odot}}}$} &\colhead{$\log{L_{5100}/\rm{erg \cdot s^{-1}}}$} & \colhead{$\dot{m}_{edd}$} \\
\colhead{(1)} & \colhead{(2)} & \colhead{(3)} &\colhead{(4)} & \colhead{(5)} 
}
\startdata
NGC 5548 & 0.017 & 7.72 & 43.36 & 0.05 \\
NGC 2617 & 0.014 & 7.51 & 43.63 & 0.01 \\
NGC 4151 & 0.003 & 7.56 & 42.32 & 0.07 \\
Mrk 142  & 0.045 & 6.23 & 43.31 & 0.82 \\
NGC 4593 & 0.008 & 6.88 & 42.87 & 0.08 \\
Mrk 509  & 0.034 & 8.05 & 44.15 & 0.05 \\
\enddata
\end{deluxetable}
\label{prop}
\vspace{-1.5cm} 

\section{Result}

\subsection{The ICCF-Cut Results}

\begin{figure*}[t]
  \centering
  \includegraphics[width=0.92\textwidth]{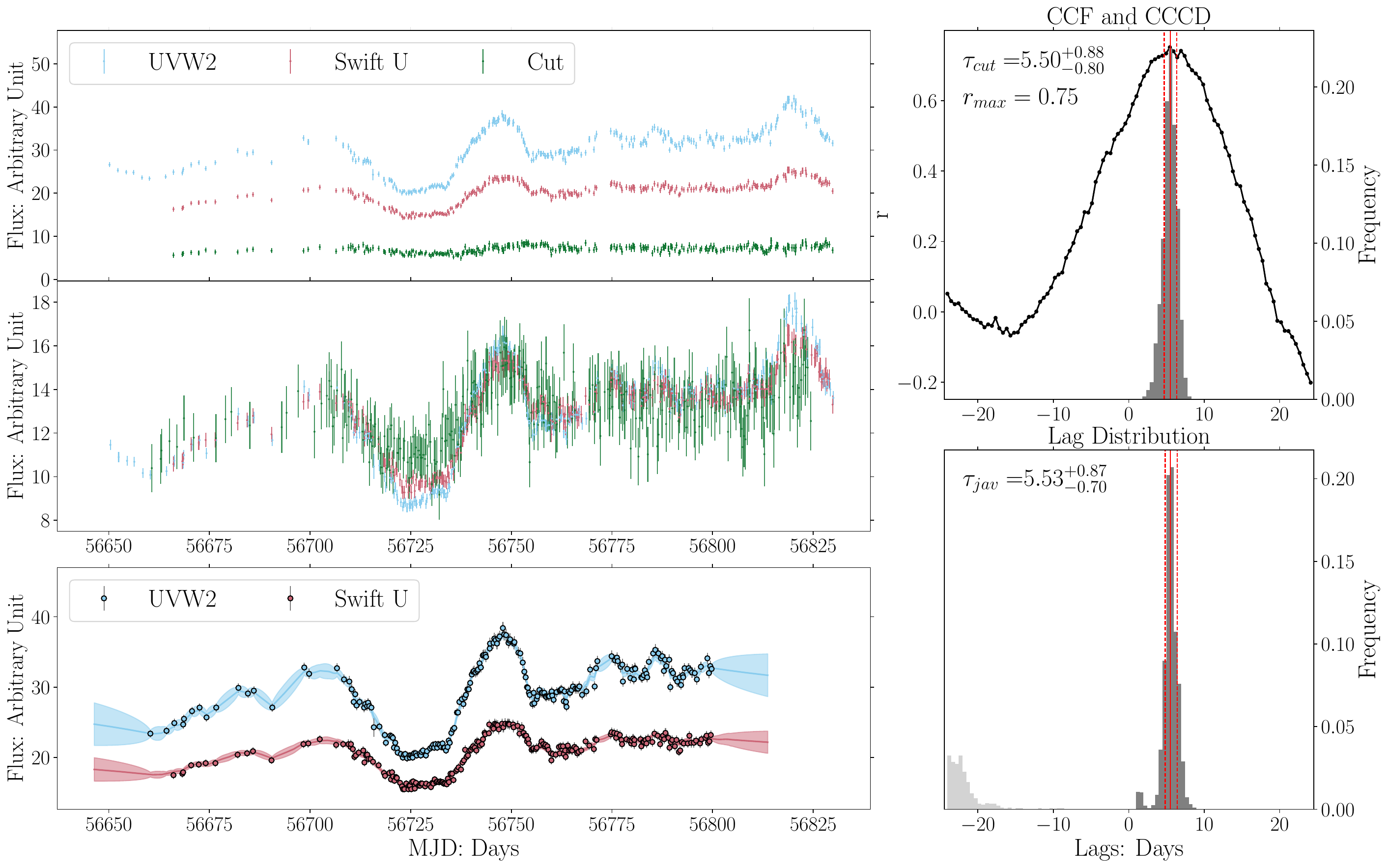}
  \caption{The ICCF-Cut and JAVELIN Pmap Model results for NGC 5548. The top-left panel shows the ICCF-Cut light curves, including the observed UVW2 (blue), observed Swift U (red) and the cut (green; i.e., the observed Swift U band light curve minus the predicted disk emission light curve in that band) light curves. The upper subpanel displays original light curves with host galaxy flux subtracted, while the lower subpanel exhibits light curves scaled to the predicted disk flux in the Swift U band and shifted by their respective lags. The bottom-left panel shows the observed light curves and the JAVELIN fittings in the UVW2 and Swift U bands. The top-right panel presents the cross-correlation function (CCF) and cross-correlation centroid distribution (CCCD) between the UVW2 and cut light curves. The bottom-right panel shows the posterior distribution of lags given by the JAVELIN Pmap Model. The masked data in JAVELIN results are denoted by light grey. The maximum of the CCF, the lag estimation of the ICCF-Cut method and the JAVELIN Pmap Model, $r_{max}$, $\tau_{cut}$ and $\tau_{jav}$ respectively, can be found in the corresponding panels. All lags presented here are in the rest frame.}
  \label{lc_all_ngc5548}
\end{figure*}

\begin{figure*}[]
  \centering
  \includegraphics[width=0.92\textwidth]{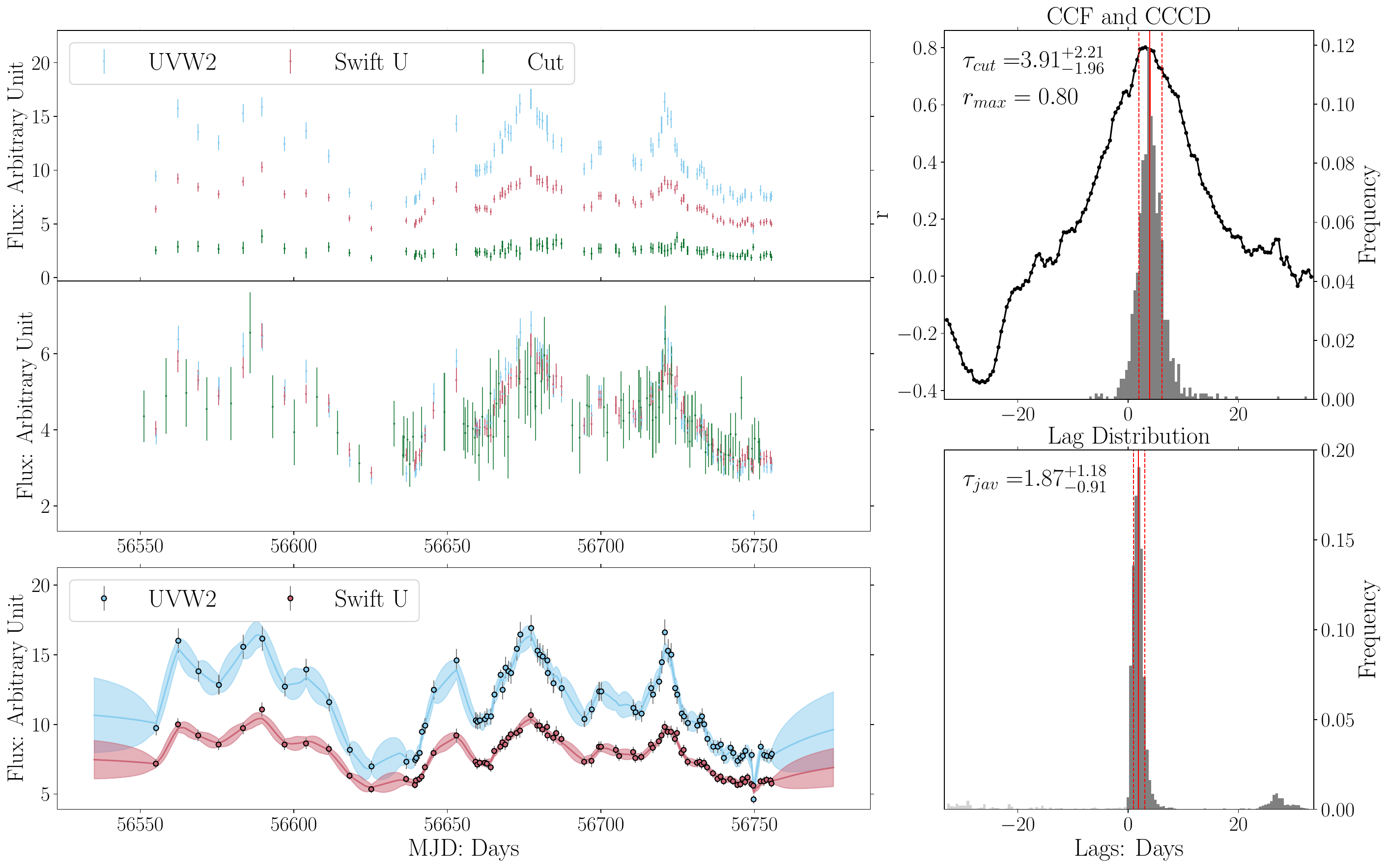}
  \caption{Same as Figure \ref{lc_all_ngc5548} but for NGC 2617.}
  \label{lc_all_ngc2617}
\end{figure*}

\begin{figure*}[]
  \centering
  \includegraphics[width=0.92\textwidth]{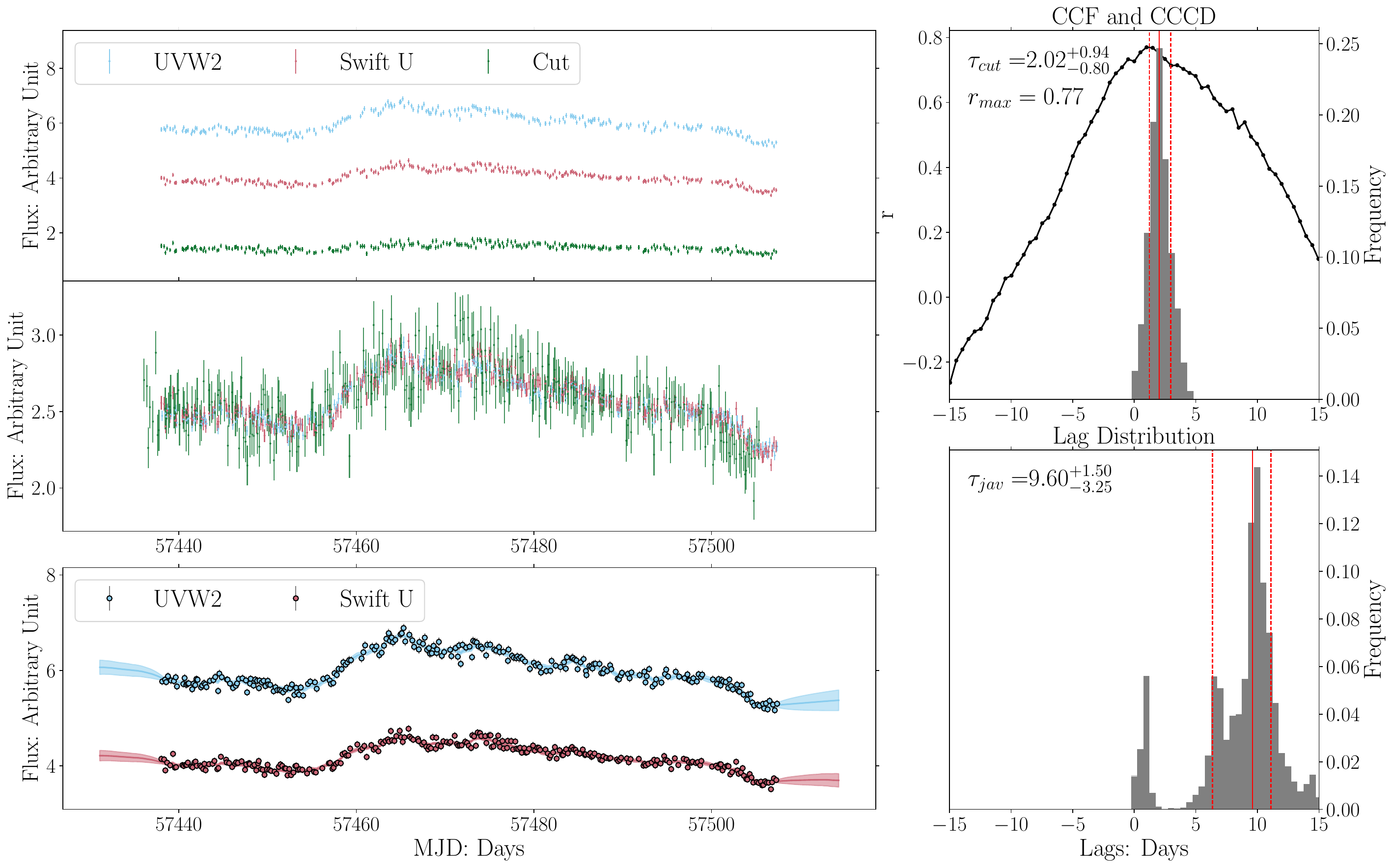}
  \caption{Same as Figure \ref{lc_all_ngc5548} but for NGC 4151.}
  \label{lc_all_ngc4151}
\end{figure*}

\begin{figure*}[]
  \centering
  \includegraphics[width=0.92\textwidth]{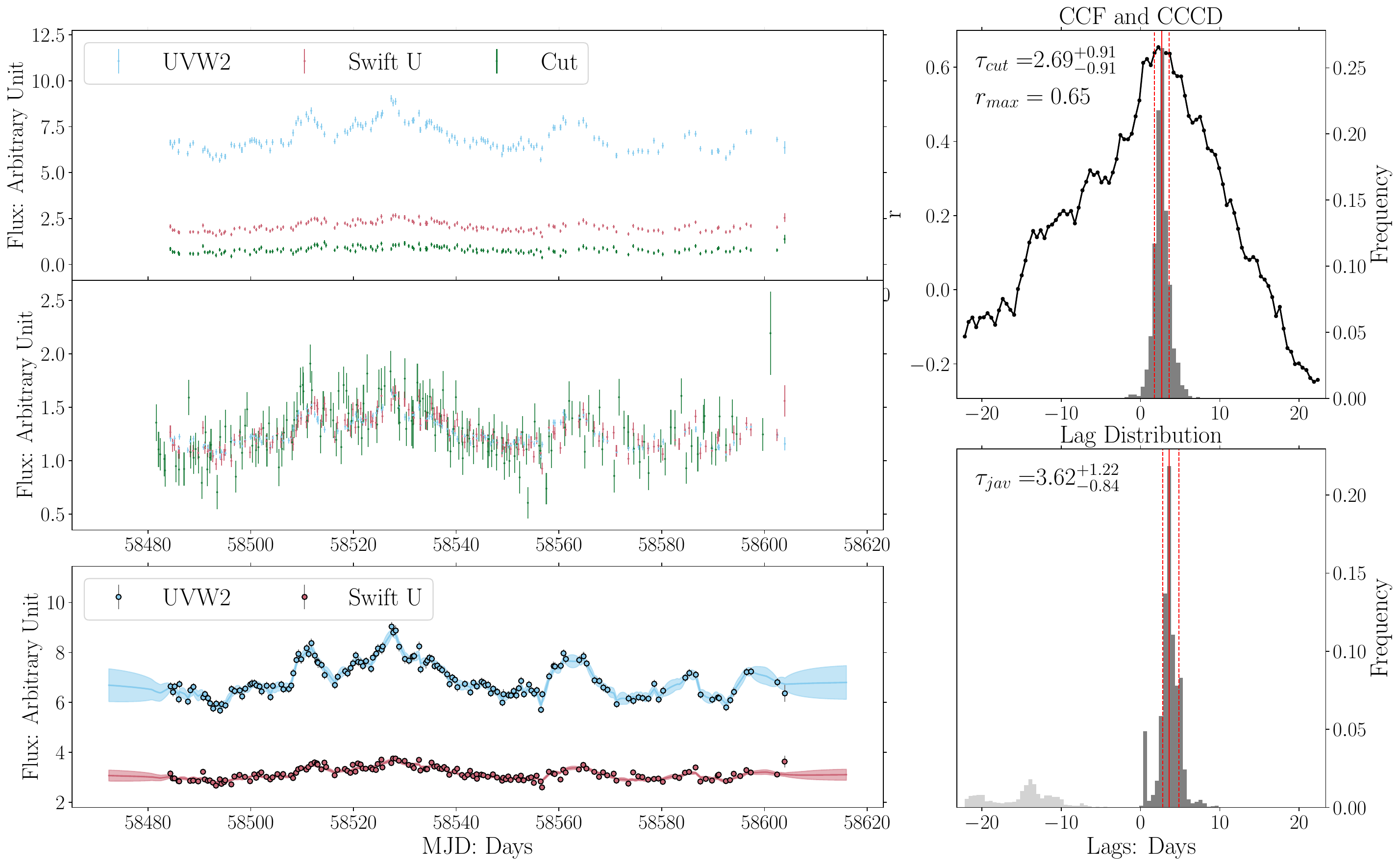}
  \caption{Same as Figure \ref{lc_all_ngc5548} but for Mrk 142.}
  \label{lc_all_mrk142}
\end{figure*}

\begin{figure*}[]
  \centering
  \includegraphics[width=0.92\textwidth]{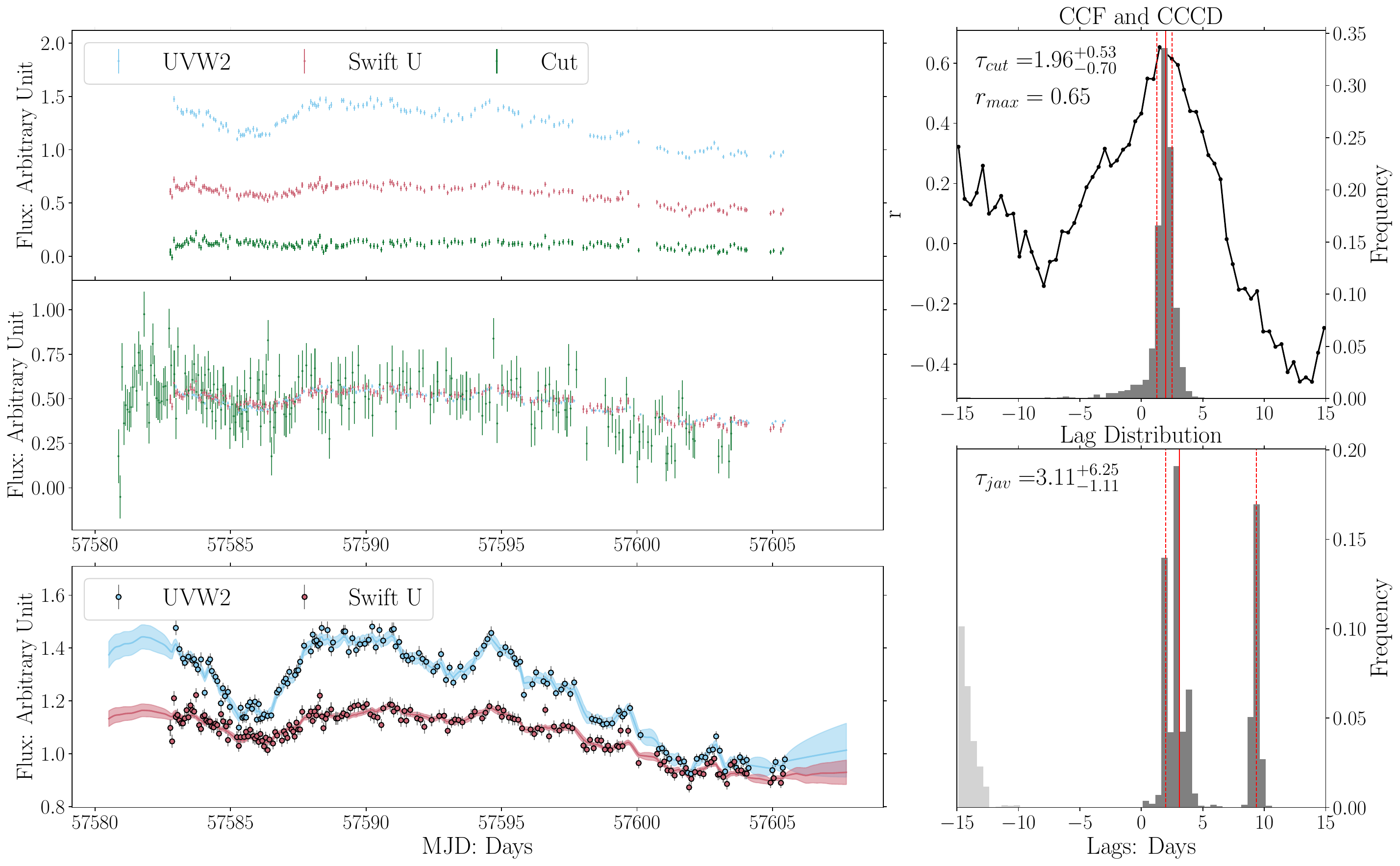}
  \caption{Same as Figure \ref{lc_all_ngc5548} but for NGC 4593.}
  \label{lc_all_ngc4593}
\end{figure*}

\begin{figure*}[]
\centering
  \includegraphics[width=0.92\textwidth]{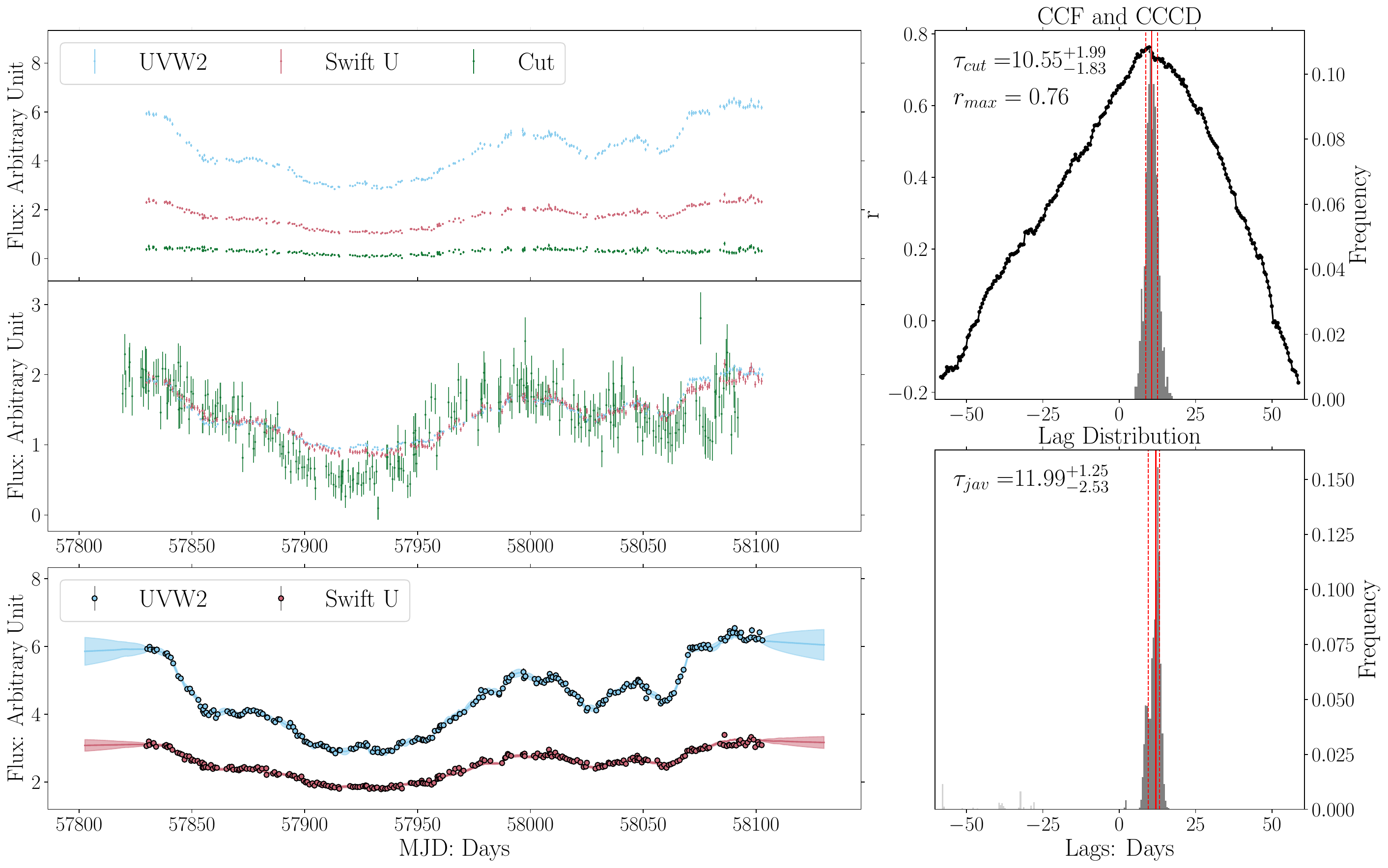}
  \caption{Same as Figure \ref{lc_all_ngc5548} but for Mrk 509.}
  \label{lc_all_mrk509}
\end{figure*}

We apply the ICCF-Cut method for continuum reverberation mapping to all targets in our sample. First, we follow the strategy mentioned in Section 2.1, estimating the host flux, disk ratio and disk lag for each target. Then, we derive the disk component from the Swift U band. For convenience, we refer the UVW2 band light curve to the``driving" light curve because it is in the band with the shortest wavelength of our sample and the pure disk component based on our assumptions. We call the light curve in the Swift U band as the ``target” light curve, and the derived light curve for the possible diffuse continuum (i.e., the observed Swift U band light curve minus the predicted disk emission light curve in the Swift U band) as the ``cut" light curve. Finally, we use the ICCF method to measure lags between the driving and cut light curves. 

Some previous works predicted that the diffuse continuum lag is about $\tau_{dc} = 0.5r(\rm{H\beta})$ \citep{2020MNRAS.494.1611N}. So we set the lag search range to be $-3 \tau_{dc} \sim 3\tau_{dc}$ and $r(\rm{H\beta})$ is estimated by the $R-L$ relation \citep{2013ApJ...767..149B}. If $3\tau_{dc}$ is less than $15$ days, we set the lag search range from -15 to 15 days to avoid missing longer lags. We increase the lag search range from 15 days to 3/4 times of the whole duration of each observations for those targets and find the resulting changes remaining minimal compared to the 15-day range, so we consider the 15-day range to be an appropriate limit. Then, we generate a time grid by $0.5$ days and use it in ICCF.

We found a good correlation and larger lag for each target in our sample. Our results for the diffuse continuum lag measurement are shown in Figures \ref{lc_all_ngc5548} - \ref{lc_all_mrk509} and Table \ref{cut_result}. The maximum correlation coefficient $r$ is larger than $0.6$ for every target in our sample, which means that the variabilities of driving light curves are similar to the diffuse continuum light curves. Moreover, the new lag $\tau_{cut}$ is significantly larger than the original lag $\tau_{ori}$. $\tau_{ori}$ is the lag calculated without any cutting process. In Figures \ref{lc_all_ngc5548}-\ref{lc_all_mrk509}, we also scale all light curves to the disk flux and shift them by the corresponding lags. It is obvious that the driving, target and cut light curves are consistent after the scaling and shifting. All evidence indicates that the new lag is not only larger but also reliable. This result is consistent with the picture that the central emission drives the Balmer continuum in the BLR.

We also compare our results with the measured $\rm{H\beta}$ lags. We found that for most of the targets, the lag of a outer component in Swift U band is consistent with predicted Balmer continuum lag, roughly $0.5 \tau(\rm{H \beta})$ (see the comparison in Table \ref{cut_result}). The difference may be caused by uncertainties, such as the different disk lags and other possible contaminations. We will discuss those uncertainties of the parameter choices and check their impacts on lag measurements in the next section via simulations. Here one exception is Mrk 509. Our result is far smaller than the predicted one. But in general, we found a larger and reliable lag that is consistent with the predicted Balmer continuum lag approximately, which provides a more direct evidence on the existence of an outer component in the broad-band photometric light curve. This component exhibits a good correlation with the central disk emission and is located in close proximity to the Balmer continuum.

Although the ICCF-Cut results have reached agreement with the predictions approximately, the ICCF-Cut method for the continuum reverberation mapping may suffer from uncertainties of its parameters. Firstly, the disk lag and disk ratio are both predicted by models. The disk lags used here are predicted by the standard thin disk model. We can find some outer components that can influence our lag measurement by assuming this disk model is correct, but we cannot ensure that this disk model itself is correct. Another disk model called Novikov-Thorne accretion disk has been discussed recently \citep{2021ApJ...907...20K,2021MNRAS.503.4163K,2023MNRAS.526..138K}. This model not only considers a lamp-post model and a thermal reverberation mapping, but also further considers more elaborate mechanisms and parameters, such as the relativistic effects and disk reflection. They could fit the lag-spectra well when excluding the u/U band, which means there is no need for BLR emissions such as diffuse continuum to explain the larger lags in other bands. The u/U band excess may also be caused by the Balmer continuum because it is very strong. This result agreed with our result to some extent.

Additionally, the disk ratio is predicted by CLOUDY and a specific BLR model. The model also depends on a parameter - covering factor $c_f$. We use the average $c_f = 0.2$ for most targets because the lag-spectra predicted by it is similar to the observational lags in general. Additional evidence could prove that the ratios of $\rm{H\beta}$ lags and the non-cut continuum lags from a sample of 21 AGNs are generally consistent with the prediction assuming $c_f = 0.2$ \citep{2023ApJ...948L..23W}. But for NGC 4593 and Mrk 509, we found that if we use $c_f = 0.2$, the predicted lags are too big, so we use $c_f = 0.1$ instead, as in \citet{2022MNRAS.509.2637N}. 

\begin{figure*}[!htb]
  \includegraphics[width=1.0\linewidth]{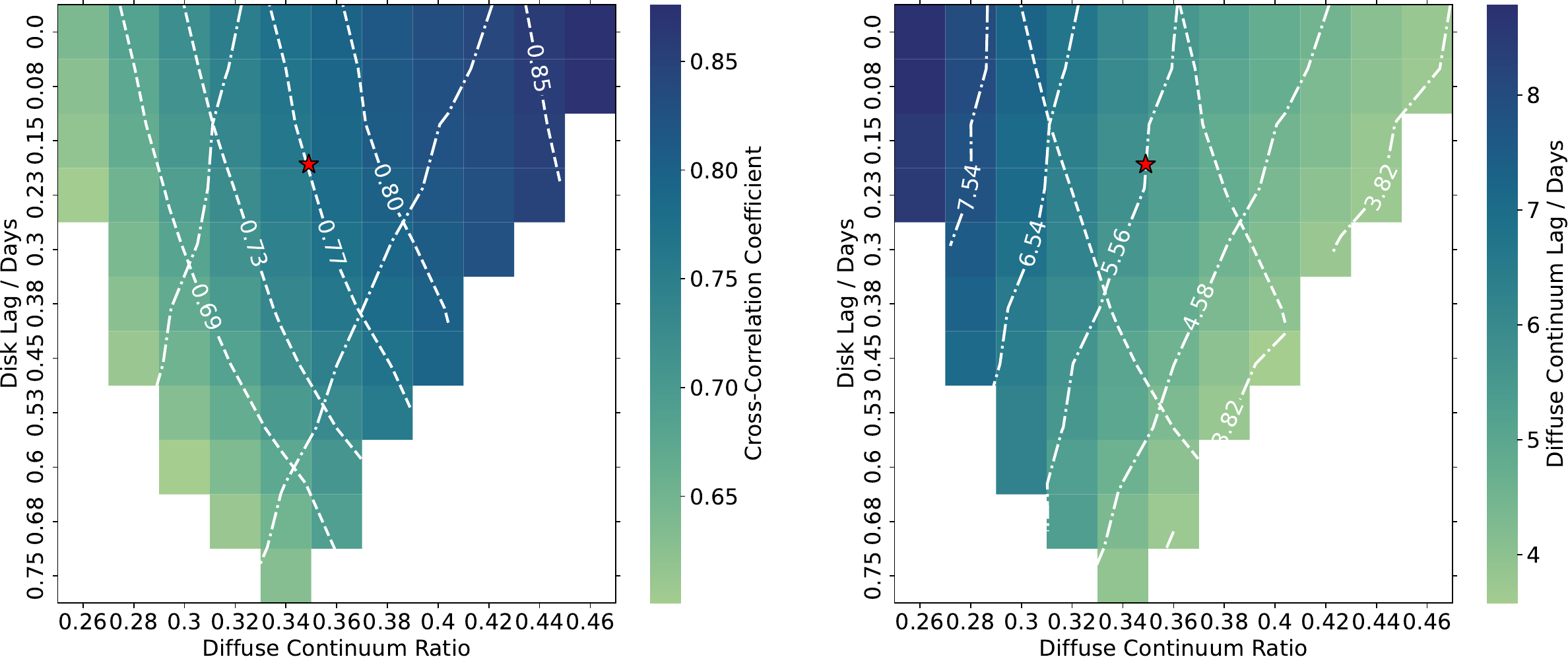}
  \caption{Simulation results for NGC 5548 about the influences of the disk lag $\tau_{disk}$ and the diffuse continuum ratio $p_{dc}$. Only good results are displayed here for clarity. The left panel is the heat map of the maximum cross-correlation coefficient $r_{max}$. The right panel is the heat map of the resulting diffuse continuum lag $\tau_{dc}$. The dashed lines in both panels represent the contours of $r_{max}$, while the dotted dashed lines in both panels represent the contours of $\tau_{dc}$. The values of the contour lines are marked. The red stars in two panels show the ICCF-Cut results with $\tau_{disk}$ predicted by the standard thin disk model and $p_{dc}$ predicted by the CLOUDY simulations. }
  \label{sim}
\end{figure*}

To fully understand the influence of disk lags and disk ratios, we perform a simple simulation. First, we generate a grid of disk lags, $\tau_{disk}$ and diffuse continuum ratio, $p_{dc}$ (or disk ratio $1-p_{dc}$).  We took 20 $\tau_{disk}$  among $[0,\tau_{obs}]$. The upper limit is $\tau_{obs}$ because we think the disk lag is the contribution of inner disk emission and shouldn't be larger than the total observed lag. The range of $p_{dc}$ is [0.1,0.5] days, and we take a point every 0.02 days. Then, we repeat our method to each point in $(p_{dc},\tau_{obs})$ and get a new lag measurement and the new cross-correlation coefficient. We regard the result that has a new cross-correlation coefficient $r_{max}>0.6$ and a new lag $\tau_{dc} > 3\tau_{obs}$ as a good result. We require a high $r$ value because we expect the diffuse continuum can be correlated to the central disk emission well, which is indicated by the $r$ coefficient. But we still need a larger lag because the resulting component is expected to be an outer component in our assumption. The highest $r$ value can always be obtained without any cutting process due to disk components. In Figure \ref{sim}, we select the good results for NGC 5548 and the heat plots for $r_{max}$ and $\tau_{dc}$. The contours of 5\% and 10\% errors of $r_{max}$ and $1 \sigma$ errors of $\tau_{dc}$ are displayed in the panel of the maximum cross-correlation coefficient, and the contours of 5\% errors of $r_{max}$ and $1 \sigma$ and $2 \sigma$ errors are displayed in the panel of the diffuse continuum lag. In general, if $p_{dc}$ is higher and $\tau_{disk}$ is smaller, the maximum of the cross-correlation coefficient $r$ will be higher; if $p_{dc}$ is lower and $\tau_{disk}$ is smaller, $\tau_{dc}$ will be larger. We note that the results for other targets are very similar, so we only show one example here. The trend shows that the appropriate $p_{dc}$ and $\tau_{disk}$ values are needed to secure a high $r_{max}$ as well as larger $\tau_{disk}$ values, though the result is not bad around the parameters given by the thin disk models and CLOUDY simulations. The simulation results suggest that a good diffuse continuum lag measurement and its consistency with the predicted Balmer continuum lag are not obtained by accident. We must choose the correct $(p_{dc},\tau_{obs})$, and our choices based on specific models are reasonable, but further constraints may still be needed. 

\begin{deluxetable*}{ccccccccc}[t]
\centering
\tablecaption{ICCF-Cut and JAVELIN Results in the Swift U band}
\tablehead{
\colhead{Name} & \colhead{$p_{dc}$} & \colhead{$r_{max}$} & \colhead{$\tau_{cut}$/days} & \colhead{$\tau_{jav}$/days} & \colhead{$\tau_{ori}$/days} & \colhead{$\tau_{dc}$/days} & \colhead{Flag}\\
\colhead{(1)} & \colhead{(2)} & \colhead{(3)} &\colhead{(4)} & \colhead{(5)} & \colhead{(6)} & \colhead{(7)} &  \colhead{(8)}
}
\startdata
NGC 5548 & 0.35 & 0.75 & $5.50^{+0.88}_{-0.80}$ & $5.53_{-0.70}^{+0.87}$ & $1.17^{+0.24}_{-0.24}$ & $5.40^{+0.19}_{-0.19}$ & 1\\
NGC 2617 & 0.37 & 0.80 & $3.92^{+2.21}_{-1.96}$ & $1.87_{-0.91}^{+1.18}$ & $0.90^{+0.51}_{-0.56}$ & $2.15^{+0.55}_{-0.70}$ & 1\\
NGC 4151 & 0.36 & 0.77 & $2.02^{+0.94}_{-0.80}$ & $9.60_{-3.25}^{+1.50}$ & $0.68^{+0.24}_{-0.24}$ & $3.41^{+0.24}_{-0.29}$ & 0\\
Mrk 142  & 0.39 & 0.65 & $2.69^{+0.91}_{-0.91}$ & $3.62_{-0.84}^{+1.22}$ & $1.04^{+0.11}_{-0.10}$ & $3.95^{+0.60}_{-0.55}$ & 1\\
NGC 4593 & 0.18 & 0.65 & $1.96^{+0.53}_{-0.70}$ & $3.11_{-1.11}^{+6.25}$ & $0.33^{+0.11}_{-0.11}$ & $1.87^{+0.36}_{-0.36}$ & 0\\
Mrk 509  & 0.18 & 0.77 & $10.55^{+1.99}_{-1.83}$ & $11.09_{-2.53}^{+1.25}$ & $2.54^{+0.56}_{-0.57}$ &  $39.80^{+3.05}_{-2.70}$ & 1\\
\enddata
\tablecomments{Column 1: Target Names. Column 2: Diffuse Continuum Ratios $p_{dc}$ in the Swift U band predicted by CLOUDY. Column 3: The Maximum Cross-Correlation Coefficient $r_{max}$. Column 4 and Column 5: ICCF-Cut and JAVELIN Pmap Model Lags, $\tau_{cut}$ and $\tau_{jav}$ respectively. Column 6: Original Lag, $\tau_{ori}$, from the corresponding papers. Column 7: Predicted Balmer Continuum Lags $\tau_{dc} = 0.5\tau_{\rm{H\beta}}$. The errors are transmitted by the observed $\rm{H \beta}$ lags. Column 8: Core Sample Flags. If the flag equals to 1, it means this target has consistent ICCF-Cut and JAVELIN Pmap Model lags, and it is categorized as core sample. All lags presented here are in the rest frame. The $r_{max}$, $\tau_{cut}$ and $\tau_{jav}$ are measured between the driving (UVW2 observed) light curves and cut light curves, the ones we subtract the predicted disk components from the Swift U bands. }
\end{deluxetable*}
\label{cut_result}
\vspace{-1.2cm}

\subsection{The JAVELIN Pmap Results}

After using ICCF-Cut, we also apply the JAVELIN Pmap Model to each target in our sample. We set the lag search range from $-3\tau_{dc}$ to $3\tau_{dc}$ days (or -15 to 15 days if $3\tau_{dc}<15$). For MCMC parameters, we use $n_{chain} = n_{walkers} = n_{burn} = 200$ to ensure a sufficient sampling. The posterior distribution of our sample given by the Pmap Model can be found in Figures \ref{lc_all_ngc5548}-\ref{lc_all_mrk509}. We find some targets have negative peaks in their lag distribution. Most of those peaks are located on the negative edge of the lag distribution and correspond to a very low diffuse continuum ratio. The negative peaks also appear in other RM projects and are thought to be mainly caused by noise or sampling properties because a negative lag is unphysical in the reverberation mapping framework \citep[e.g.,][]{2017ApJ...851...21G}. Therefore, we think that those negative peaks are unphysical and suspicious, and mask them when estimating the final JAVELIN Pmap Model lags. The final results of JAVELIN can be found in Table \ref{cut_result}. 

The lags given by the Pmap model are similar to that of ICCF-Cut for NGC 5548, NGC 2617, Mrk 142 and Mrk 509, while not consistent for NGC 4593 and NGC 4151. The centroid lags $\tau_{cut}$ given by ICCF-Cut and the median lags given by JAVELIN for the four targets have similar values and are consistent with each other within the error bars, which indicates the reliability of their ICCF-Cut lags.

JAVELIN lag $\tau_{jav}$ of NGC 4151 is much larger than the ICCF-Cut result, while the lag distribution of NGC 4593 has multiple strong peaks, which makes it suspicious to be a true estimation, even though the JAVELIN lag could reach consistency within the error bar. We note that these two targets have short observation baselines. The monitoring duration is 69.3 days for NGC 4151 and 22.6 days for NGC 4593 \citep{2019ApJ...870..123E} while the duration of other targets is more than 100 days. This may explain why those two targets cannot have good and consistent results, and the predicted lags may not be convincing. When the observational period is short, shifting the light curve by a long lag will have fewer overlapping parts and be more likely to result in lower correlations or poorer lag distributions. Despite this, additional peaks in the lag distributions can occasionally emerge coincidentally, such as those long-lag peaks ($\sim 10$ days) in the lag distributions of NGC 4151 and NGC 4593. One possible explanation for those strong peaks is the existence of similar local features in their light curves. Those features may mismatch each other and contribute to those peaks with longer lags. 

In conclusion, the JAVELIN Pmap Model may be feasible in recovering the diffuse continuum lag due to the consistency with the lag results for 4 of 6 targets in our sample. However, we note again that the JAVELIN method relies on the DRW model and a simple transfer function, which still needs to be tested in the future. Also, there are often some outliers caused by the current quality of data. Therefore, the JAVELIN Pmap Model would be better to be used to evaluate the ICCF-Cut lag measurements. We select a core sample from our parent sample of 6 AGNs, including 4 targets, NGC 5548, NGC 2617, Mrk 142 and Mrk 509, because they have consistent lag measurements by both the ICCF-Cut method and the JAVELIN Pmap Model. We think the lags in the core sample are more reliable. 

\section{Discussion}

\subsection{Extensions to Other Bands}

According to some previous works \citep[e.g.,][]{2018MNRAS.481..533L,2019MNRAS.489.5284K,2022MNRAS.509.2637N}, diffuse continuum spread out the whole UV/optical spectrum and contribute to the observed lags. So, we want to check if we could find similar outer components in other bands. We expand the ICCF-Cut method to other bands of NGC 5548. There are two reasons why the extension is only applied to NGC 5548. First, we have data from 14 bands for NGC 5548, which enables our extension. Second, NGC 5548 has more accurate host galaxy flux measurement by the image-subtraction method instead of the estimation by the flux-flux analysis. We still use the UVW2 band as the driving light curve because it has less contamination by the diffuse continuum. The only thing to be changed is to switch all the values for the Swift U band to other bands (target bands hereafter). At last, we perform the ICCF analyses between the UVW2 light curves and the cut light curves derived from the target bands.

\begin{figure*}[]
  \includegraphics[width=1.0\linewidth]{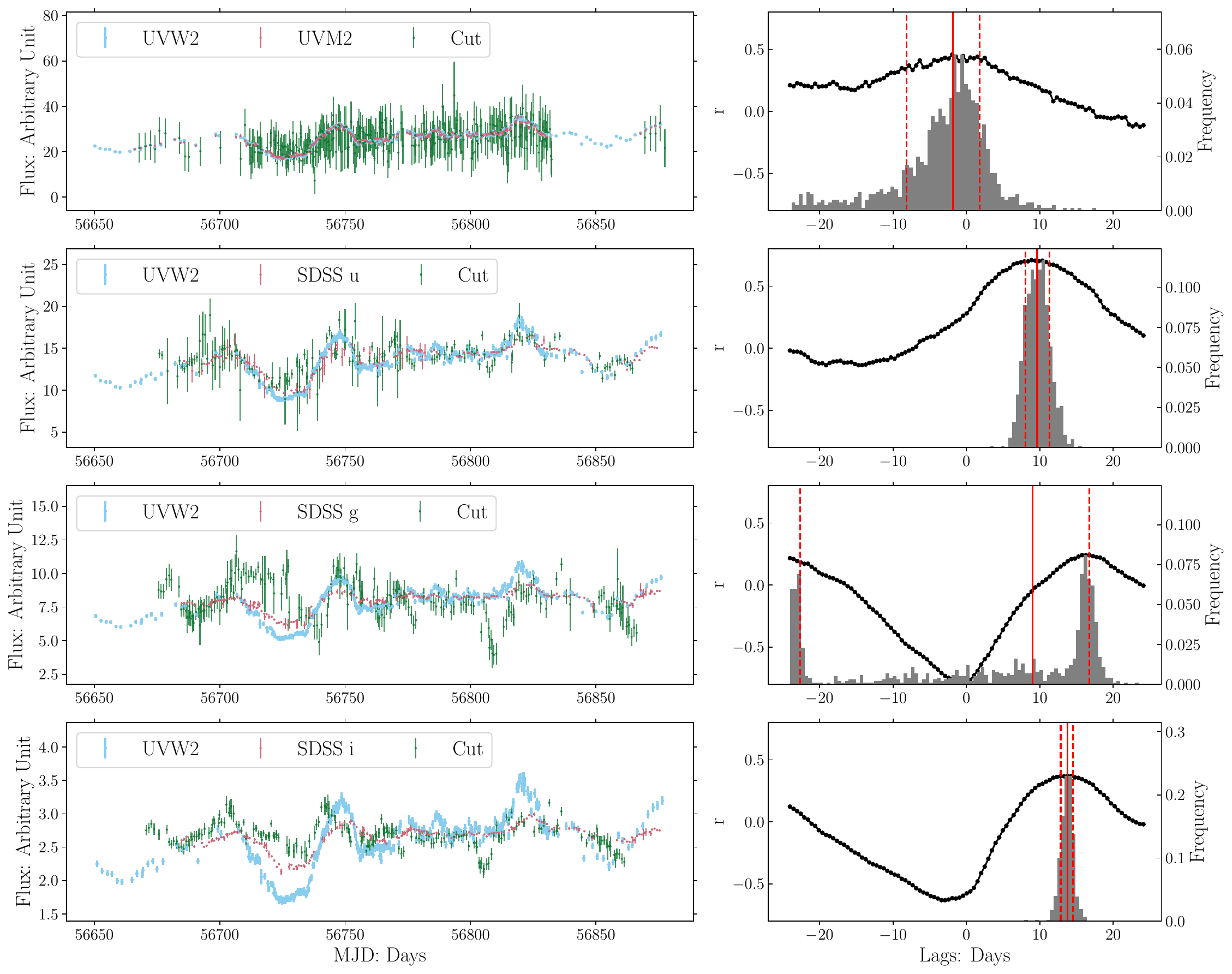}
  \caption{Examples of the ICCF-Cut results for NGC 5548 in other bands. From top to bottom, the panels show results for UVM2, SDSS u and SDSS g bands, respectively. The panels on the left are the driving light curves, corresponding target light curves and cut light curves. They are all scaled to the same scale and shifted by lags. The panels on the right are ICCF results. The black curves show the CCF while the grey histograms show the CCCD. The red solid lines show the centroid lags and the red dashed lines show the lag uncertainties. }
  \label{all_band}
\end{figure*}

By expanding our approach to all other bands of NGC 5548, we find good correlation in the SDSS u band. On the contrary, the correlation is bad in other bands. SDSS u band has a similar wavelength coverage to the Swift U band and involves a strong Balmer continuum in it. The new lag is $9.49^{+1.85}_{-1.64}$ days in the rest frame and $r_{max} = 0.69$. In other bands, the maximum correlation coefficient is very low, so the result is not reliable even if larger lags can be found in some bands. We display the results for four different bands in Figure \ref{all_band}. Results in other bands are similar and show no correlations. We will discuss the possible explanations in Section 5.2. This result means that our method can extract an outer component which is out of the accretion disk for the u/U band where the diffuse continuum is very strong, but it does not show the contributions of other outer components to the continuum reverberation mapping from other bands.

Above all, we found an outer component beyond the accretion disk in the Swift U band for 6 AGNs. It has a good correlation to the central disk continuum and also provides a larger lag than the disk lag. The result illustrates the contribution of an outer component to the continuum reverberation mapping from the light curves directly. Moreover, this component covers the wavelength of the Balmer continuum and also has a similar lag of it. But we noticed that: 1) The lags after cutting, $\tau_{cut}$, are not consistent with the prediction exactly, especially for Mrk 509; 2) In a detailed investigation of NGC 5548, we only find similar result in SDSS u band where Balmer continuum is also very strong, and in other bands, a good correlation and a larger lag are not observed. We will discuss some possible explanations in the following sections.

\subsection{Paschen Continuum and Emission Lines}

A larger lag and a good correlation are expected to be found after subtracting the predicted disk components in other bands with a significant diffuse continuum ratio. However, our results show that the correlation and the shape of the CCF are not good in bands other than Swift U and SDSS u bands, even for the SDSS $i$ band, where the Paschen continuum is located for NGC 5548. From the CLOUDY simulation, the diffuse continuum ratio in $i$ band is $p_{dc} = 0.46$, and the predicted Paschen lag by \citet{2020MNRAS.494.1611N} is also about $0.5\tau(\rm{H\beta})$. However, in our result, although the $i$ band light curves have some similar features in Figure \ref{all_band}, and the cross-correlation coefficient seems a little bit higher than those in other bands like the SDSS g band, the $r_{max}$ is still very low, and the shape of the CCF is strange. 

There are two main explanations for this result. Firstly, a standard thin disk model may be too simplistic. If we input another disk SED, the diffuse continuum ratio in Figure \ref{dc_ratio} and the predicted disk lags will change, which will alter the predicted disk component in the ``target" bands (the bands used to extract the diffuse continuum). In \citet{2021ApJ...907...20K, 2021MNRAS.503.4163K}, they constructed a standard Novikov–Thorne accretion disk around a rotating black hole and found that the predicted lags are consistent with the observed lags when excluding the u/U band lags. This is similar to our results, which show larger lags and good correlations in the u/U bands but no good correlations in other bands. A lag change in the U band will not change the result too much (see a simple test in Section 5.3). However, since we input a standard thin disk SED for CLOUDY simulations, the wavelength distribution of the predicted diffuse continuum ratio in all bands could also change after the input SED change. Consequently, further investigation into the influence of different disk SEDs is needed in future works.

Secondly, some other components in the band could also influence the results when those components are compatible with the disk continuum or the diffuse continuum. Some line emissions will also contribute to the total flux. For example, $\rm{C \ III]}$ locates in the UVW2 band; $\rm{H\beta}$ locates in the g band, and $\rm{H\alpha}$ locates in the r/R band; $\rm{Fe \ II}$ emission is a broad contribution, blended between $\sim 2000-4000 \rm{\mathring{A}}$ \citep{1983ApJ...275..445N,1985ApJ...288...94W}; the high-order Paschen continuum and the dust emission plays an important role in the SDSS $i$ band. Those components may also contribute to the total photometric lag, but we did not remove them from the total light curves in the ICCF-Cut method. 

For the driving light curves in the UVW2 band, the continuum component is extremely significant, and the typical contamination of $\rm{CIII]}$ is less than 5\% based on the composite quasar spectra \citep{2001AJ....122..549V}, so the line emission only have minimal impact on the ICCF-Cut results. For the ``target" light curves, the relative flux of the disk and diffuse continuum compared to other components can influence our results. Supposing that the disk and diffuse continuum flux are weaker or comparable with other components, the basic assumption about only two main variable components in each band is incorrect, resulting in overestimating the disk continuum ratio. If we still assume the existence of the contribution from the components other than the disk and diffuse continuum, the total light curves may be written as
\begin{eqnarray}
\label{eq_tot_j}
L_{j} (t) = L_{j,disk}(t) + L_{j, dc} (t) + L_{j, other} (t),
\end{eqnarray}
where $L_{j} (t)$, $ L_{j,disk}(t) $, $ L_{j,dc} (t) $, $ L_{j,other} (t) $ represent the total, disk, diffuse continuum and other possible emissions. The footnote $j$ represents the ``target" band $j$. The parameter $\alpha$, which is used to transfer the flux from the UVW2 band to the other band, will change to
\begin{equation}
\begin{split}
\alpha &= \frac{L_{j,disk}(t)}{L_{W2,disk}(t-\tau_{disk})} \\
&= \frac{L_{j,disk}}{L_{j,disk} + L_{j,dc}+ L_{j,other}} \cdot \frac{L_{j}}{L_{W2}}.
\end{split}
\label{eq_trans_line}
\end{equation}
In the simulation works \citep[e.g.,][]{2018MNRAS.481..533L,2019MNRAS.489.5284K,2022MNRAS.509.2637N}, researchers used $p_{j,disk} = 1 - p_{j,dc} = L_{j,disk}/(L_{j,disk}+L_{j,dc})$ to simply predict the observed lags. This simple estimation could be feasible if only the disk and diffuse continuum can provide observable lags. However, when transferring the flux, we should use the disk ratio over the total flux. In the Swift U band, the disk continuum $L_{U,disk}$ and the diffuse continuum $L_{U,dc}$ are very strong, so ignoring the emission of other components $L_{U,other}$ may not influence our results too much even after cutting the disk component off the Swift U band. The strong correlation further supports this assertion. However, in other bands, we may overestimate the first term of the transfer parameter $\alpha$ due to other emissions so that we will remove more disk flux than the true value. Since the disk lag is smaller and close to 0, the CCF between two disk components will be located near 0, so the negative peak of the CCF around 0 in Figure \ref{all_band} could be caused by the over removal of the disk component. For example, some works \citep[e.g., ][]{1982ApJ...254...22M, 2022ApJ...927...60G} states that the spectral break at the Paschen jump can be offset by the excess flux of high-order Paschen lines, which means that the flux of high-order Paschen lines is similar to that of the Paschen continuum. Additionally, the dust emission could also exit in the SDSS $i$ band and affect the disk continuum fraction over the total flux \citep{2022ApJ...940...20G, 2022MNRAS.509.2637N}. Hence, we are likely to overestimate the disk ratio as a result of ignoring other emissions in the $i$ band and get a negative peak in the CCF. This is also a possible explanation for why the result is good in the Swift U band but bad in other bands.

\subsection{Other Uncertainties}

There are other uncertainties that we should also consider. Firstly, the host galaxy flux also contributes to the broad-band photometric light curves. Most host flux is estimated by flux-flux analysis rather than the image-subtraction method, which is more accurate. As mentioned above, some works have found that the host galaxy flux estimated by the flux-flux analysis is larger than that given by the image-subtraction \citep{2017ApJ...835...65S}. Due to these uncertainties, the final lags may not be consistent with the Balmer continuum lag predictions exactly. 

According to Equation (\ref{eq_trans_cut}), we transfer the disk flux from the UVW2 band to the Swift U band by a scaling factor $\alpha = (1-p_{dc})\times \rm{Median}(L_{U}/L_{W2})$. If we subtract more disk flux from the U band light curve, we will underestimate the $\alpha$ value. A lower $\alpha$ value could result in remaining disk flux in the cut light curve after we get rid of the disk components in the Swift U band, which will reduce the ICCF-Cut lags $\tau_{cut}$ between the driving (UVW2) light curve and the cut light curve. If we use the flux-flux analysis result for NGC 5548, which makes the mean flux (without host galaxy contribution in the Swift U band) decreases about 15\%, the ICCF-Cut lag $\tau_{cut}$ will change to $2.97^{+0.55}_{-0.69}$, and $r_{max}=0.89$. This lag is smaller compared to our ICCF-Cut lag measured in Section 4 where $\tau_{cut}=5.50^{+0.88}_{-0.80}$ for NGC 5548, but it is still about three times larger than the original lags, $\tau_{ori} = 1.17^{+0.24}_{-0.24}$. This effect could explain the smaller ICCF-Cut lags $\tau_{cut}$ compared to the predicted diffuse continuum lags $\tau_{dc}$ for NGC 4151 and Mrk 142. Although NGC 2617 has larger $\tau_{cut}$ than the prediction, the errors of $\tau_{cut}$ are also much larger. If we have an accurate estimation of host flux for Mrk 509, the ICCF-Cut lag may increase, but we think that the huge gap between the ICCF-Cut lag $\tau_{cut}$ and predicted diffuse continuum lag $\tau_{dc}$ cannot be fully offset. For NGC 4593, $\tau_{cut}$ and $\tau_{ori}$ are similar, which means the uncertainties of the host flux may not be significant in this target. To summarize, using the flux-flux analysis method may overestimate the host flux and decrease the ICCF-Cut lags. The uncertainties could cause differences between the measured lags and the predicted lags of the diffuse continuum. However, it can still give us high correlations and larger lags compared to $\tau_{ori}$ in the Swift U band. Therefore, we expect to derive a more precise host flux in the future reverberation mapping project if we want to extract more accurate potential diffuse continuum lags by the ICCF-Cut method.

Secondly, a caveat on different choices of the disk models and the AGN ionization models should be stressed again. They can bias the result, as shown by the simulations in Section 4.1. If the disk lag is larger or the diffuse continuum ratio is lower, the diffuse continuum lag will be larger, as we illustrated in Figure \ref{sim}. However, as mentioned above, for different BLR models, CLOUDY gives a similar result around the Balmer jump, so the ratios used in the Swift U band are more reliable. Moreover, if we predicte the disk lag by the disk model in \citet{2021MNRAS.503.4163K}, the ICCF-Cut lag in the Swift U band will decrease to $2.44^{+1.53}_{-1.53}$ days in the rest frame, and the $r_{max}$ is $0.64$. This lag is smaller than that given by the standard thin disk model but is still about two times larger than the original measured lag. However, illustrated in Section 5.2, if we change the different disk model, the SED will also change and influence the CLOUDY simulation and its prediction about Balmer continuum lags. Further discussions on this issue will be present in the future work.

\section{Summary}

The main finding of this paper is:

(i) We derive a component that has a good correlation with central disk emission and a larger lag than the original lag from the Swift U band photometric light curve for 6 AGNs by a method called ICCF-Cut. The JAVELIN Pmap Model can yield comparable lags for four of six targets. 

(ii) This component covers the same wavelength range as the Balmer continuum and is consistent with the predicted Balmer continuum lag.

(iii) In a detailed analysis of NGC 5548, we only found the outer components in the Swift U and SDSS u bands where the Balmer continuum is strong. Possible explanations for not having good results in other bands could be different disk and BLR models, multiple components in the bands and uncertainties of our parameters. 

In summary, our result proves that an outer component can contribute to the continuum reverberation mapping directly and can explain the u/U band excess by the Balmer continuum well. Although current results still suffers from uncertainties in AGN parameters and models as well as other blended components in the broad-band photometric data, the results can be more accurate and robust if we have more precise measurements for relevant parameters, a better understanding of the AGN models, and the possibilities to study the contributions of Balmer continuum for more targets with UV light curves in the future. 

~

\noindent We appreciate the constructive comments and suggestions from the anonymous referee, which help improve the paper significantly. This work is supported by the National Key R\&D Program of China No. 2022YFF0503401. We acknowledge the science research grant from the China Manned Space Project with No. CMS-CSST-2021-A06. We are thankful for the support of the National Science Foundation of China (11927804 and 12133001). We thank the entire Swift team for their hard work and tireless dedication in creating the UVOT filters and executing the monitoring flawlessly. This work would not have been possible without the UVOT monitoring. We are further grateful for published data from \citet{2016ApJ...821...56F,2018ApJ...854..107F,2019ApJ...870..123E,2020ApJ...896....1C}. Their dedication in processing those data facilitated our research greatly. We acknowledge the discussions with Yuming Fu, Bing Lyu and Yuxuan Pang. Without their valuable suggestions, we can’t complete this work effectively. We also appreciate Ying Zu and Gary Ferland for their open source codes, the JAVELIN and CLOUDY \citep{2017RMxAA..53..385F}. We would like to thank them and other researchers for their contributions to the development and maintenance of those open source codes. 

\bibliography{cont-rm.bib}
\bibliographystyle{aasjournal}

\end{document}